\documentclass[a4paper,11pt]{article}
\usepackage{jinstpub} % for details on the use of the package, please see the JINST-author-manual
\usepackage{subcaption}
\title{\boldmath Radiation Tolerance Characterisation of an Indigenously 
Developed p-type Silicon Pad Sensor for Forward Calorimetry Applications}

\author[a,b,1]{Arun Kumar Yadav,\note{Corresponding author.}}
\author[a,b]{Sanjib Muhuri,}
\author[a]{Anup Kumar Sikdar,}
\author[c]{Subikash Choudhury,}
\author[d]{Sourav Mukhopadhyay,}
\author[a]{Jogender Saini,}
\author[e]{Mitul Abhangi,}
\author[e]{Ratnesh Kumar,}
\author[e,b]{Sudhirsinh Vala,}
\author[a,b]{Zubayer Ahammed}

\affiliation[a]{Variable Energy Cyclotron Centre,\\1/AF Bidhan Nagar, Kolkata 700064, India}
\affiliation[b]{Homi Bhabha National Institute, \\Anushakti Nagar, Mumbai 400094, India}
\affiliation[c]{Jadavpur University,\\ Kolkata 700032, India}
\affiliation[d]{Bhabha Atomic Research Centre,\\ Mumbai 400085, India}
\affiliation[e]{Institute for Plasma Research,\\ Gandhinagar, Ahmedabad 382428, India}

\emailAdd{ak.yadav@vecc.gov.in}

\abstract{We report on the radiation tolerance characterisation of a p-type 
silicon pad sensor indigenously designed by BARC-VECC and fabricated at 
Bharat Electronics Limited (BEL), India --- representing a significant 
step toward establishing a domestic silicon sensor manufacturing capability 
for high-energy physics applications. Single-pad test structures were 
irradiated with neutrons over a range of fluences from $\sim10^{7}$ to 
$\sim2.5\times10^{14}$~1~MeV~$n_{\mathrm{eq}}$/cm$^{2}$, spanning the 
operational regime relevant to forward calorimetry in high-luminosity 
heavy-ion collider experiments. Post-irradiation performance was 
characterised through systematic measurement of leakage current evolution 
and calorimetric response as functions of accumulated neutron fluence. 
A single-exponential annealing model is introduced to describe the 
time dependence of leakage current within the observation window, and 
the current-related damage constant $\alpha$ is extracted and compared 
with the RD48 reference value. The results demonstrate that the sensors 
survive the target fluence with measurable but recoverable degradation, 
validating the fabrication process and providing a baseline for future 
qualification of this indigenous sensor production chain.}


\arxivnumber{} % Only if you have one

\begin{document}
\maketitle
\flushbottom

\section{Introduction}
\label{sec:intro}

Contemporary high-energy physics experiments rely on a diverse suite of sensor technologies, including gas-filled sensors, scintillation-based systems, and semiconductor devices—each optimised for distinct operational regimes involving energy resolution, timing precision, and spatial granularity. Among these, silicon-based  sensors have emerged as the preferred choice for numerous applications, primarily due to their small feature size, superior energy resolution, fast response, and customisable geometrical parameters. These features render silicon sensors particularly advantageous for deployment in high-fluence radiation environments such as those encountered at the CERN Large Hadron Collider (LHC), where stringent demands on spatial accuracy and precise energy measurement prevail.

Depending on the fabrication scheme, silicon sensors are classified according to the doping of the bulk substrate. Sensors fabricated on a pentavalently doped (donor-doped) substrate are referred to as n-type, with a shallow acceptor layer subsequently implanted or diffused to form the p–n junction. Conversely, sensors constructed on a trivalently doped (acceptor-doped) substrate are termed p-type, where the junction is formed by introducing a donor-doped surface layer.
From a design standpoint, an optimally functioning silicon sensor—whether n-type or p-type—should exhibit a high breakdown voltage and minimal leakage current. Full depletion is essential for the efficient conversion of incident ionizing radiation into measurable electronic signals via  generation and collection of electron–hole pairs across the sensitive region of the sensor \cite{Casse2018}.

In practice, however, the realisation of such idealised sensor performance is impeded by the intrinsic presence of impurities and crystalline defects within the silicon bulk. While these imperfections act as trap centres that can capture mobile charge 
carriers, modern high-purity silicon sensors achieve carrier lifetimes well in 
excess of 1~ms --- orders of magnitude larger than the typical charge drift 
times of tens of nanoseconds --- such that pre-irradiation charge collection 
efficiency is not significantly affected. It is primarily after sustained 
radiation exposure that these limitations become pronounced. Extended exposure 
to ionizing radiation introduces displacement damage and defect clusters within 
the crystal lattice, leading to degradation of sensor performance over time. 
These considerations necessitate the development and qualification of 
radiation-hard silicon sensors, capable of sustaining operational integrity 
under the hostile radiation conditions typical of modern collider environments 
\cite{Li2009}.

Traditional n-type silicon sensors, although widely used because of their 
cost-effectiveness and relatively simpler fabrication, are vulnerable to 
radiation-induced type inversion \cite{Lindstrom2001,RD50}. Under high hadron 
fluence, the effective donor concentration in the n-type bulk progressively 
decreases due to donor removal and the introduction of radiation-induced 
acceptor-like defects. The full depletion voltage $V_{fd}$ initially decreases 
as the effective doping concentration is reduced, reaches a minimum near the 
type inversion point, and then rises again as radiation-induced acceptor states 
begin to dominate \cite{Moll1999}. At fluences of several $\times10^{14}$~
$n_{eq}$/cm$^2$ and beyond, $V_{fd}$ can reach 600--800~V, and LHC experiments 
typically operate in partially depleted mode at these fluences rather than 
attempting full depletion \cite{Lindstrom2001}. Once type inversion occurs, the p--n junction migrates to the side opposite 
to the readout electrodes. The signal induction is now dominated by holes 
drifting toward the backplane over the full sensor thickness, rather than 
electrons drifting the short distance to the readout electrodes. Since holes 
have lower mobility and are more susceptible to radiation-induced trapping 
than electrons, the charge collection efficiency degrades significantly, 
resulting in a deteriorated signal-to-noise ratio.

The advantages of p-type silicon sensors in radiation-hard environments have 
been well established through extensive studies by the RD50 collaboration and 
LHC experiment groups \cite{RD50,Casse2018}. P-type sensors with n-in-p 
configuration are now the baseline technology for tracking detectors in the 
ATLAS and CMS upgrades for the High Luminosity LHC, having demonstrated 
superior radiation tolerance compared to traditional n-type sensors at fluences 
up to $10^{16}$~$n_{\mathrm{eq}}$/cm$^2$ \cite{Lindstrom2001,RD50}. 
Commercial vendors such as Hamamatsu (Japan), Micron Semiconductor (UK), 
and CiS (Germany) have developed mature p-type sensor production lines 
for these applications.
What distinguishes the present work is not the sensor concept itself, but 
rather the establishment of an \textit{indigenous} p-type silicon sensor 
design and fabrication capability within India, through a collaboration 
between VECC-Kolkata, BARC-Mumbai, and Bharat Electronics Limited (BEL), 
Bengaluru. This represents a significant step toward domestic self-reliance 
in silicon detector technology for high-energy physics, and the radiation 
tolerance characterisation reported here provides the first systematic 
qualification of this indigenous production chain.

An indigenously developed segmented p-type silicon sensor array has been 
introduced~\cite{Mukhopadhyay2023}, well-suited for operation in 
high-radiation environments, exploiting the n-in-p configuration advantages 
described above.

In this article, we report on a systematic radiation tolerance study of the fabricated p-type silicon sensor. Section II outlines the sensor design and fabrication methodology. In Section III we  discuss the expected radiation damage and radiation tolerance requirements. The irradiation test setup is described in Section IV, followed by experimental results in Section V. The article is summarised in Section~VI.

\section{Development of a p-type silicon sensor}
\label{sec:design}

The R\&D effort leading to the development of the radiation-hard, segmented 
p-type silicon pad sensor array was driven by the operational requirements of 
forward calorimetry in high-luminosity heavy-ion collider experiments. The key 
design targets include a breakdown voltage exceeding 1.2~kV, full depletion 
voltage of approximately 120~V, pad capacitance of $\sim$40~pF/cm$^{2}$, and 
a leakage current of the order of a few hundred nA, while maintaining 
operational integrity up to a cumulative fluence of 
$\sim7\times10^{13}$~1~MeV~$n_{\mathrm{eq}}$/cm$^{2}$~\cite{ALICE_FoCal_TDR_2024}. 
To meet these requirements and ensure long detector lifetime under such hostile 
conditions, detailed Technology Computer-Aided Design (TCAD) simulations were 
employed to optimise both the geometrical layout and fabrication process parameters. 

The development of the p-type sensor has been carried out through a joint technical collaboration between VECC-Kolkata and BARC-Mumbai, with fabrication carried out at Bharat Electronics Limited(BEL), Bengaluru.

p-type silicon pad detectors are typically fabricated on high-resistivity p-type silicon substrates, where segmented $n^{+}$ implants form individual sensing pads. Under reverse bias, a depletion region extends into the bulk, enabling efficient collection of charge carriers generated by ionizing radiation along with superior radiation hardness compared to n-type design.

\subsection{Design challenges and optimisation}
\label{subsec:design_optimization}

The sensors in this study follow an n-in-p configuration, where arrays of n-type implants are formed on a p-type substrate. The fabrication is carried out on detector-grade, dual-side polished, prime-grade float-zone (FZ) p-type silicon wafers procured from Topsil, with a resistivity of 6--8~k$\Omega\cdot$cm and a thickness of $\sim$320~$\mu$m. Standard (non-oxygenated) FZ silicon of this class typically exhibits an interstitial-oxygen concentration on the order of $10^{15}$--$10^{16}$~cm$^{-3}$, substantially lower than oxygenated FZ or Czochralski/magnetic-Czochralski (Cz/MCz) material ($\gtrsim10^{17}$~cm$^{-3}$). Compared to n-type detectors, the design of p-type detectors is relatively more complex, as it requires at least one additional mask layer.

A major challenge is the “charge-up” effect at the Si–SiO$_2$ interface. This is mitigated by introducing an additional p-stop implantation between adjacent $n^{+}$ pad regions to ensure proper electrical isolation. In addition, along with the $n^{+}$ doped guard rings (GR), a $p^{+}$ doped edge ring (ER) is incorporated at the sensor periphery to prevent crack propagation from the scribe edge into the active region. For each single-pad test structure, the active pad has a nominal side length of 1~cm (giving the 1~cm$^{2}$ active area), separated by a 60~$\mu$m gap on each side from the innermost guard ring, which has a width of 350~$\mu$m; a 10~$\mu$m p-stop implant is placed midway within this gap, between the pad and the guard ring.

\begin{figure}
    \centering
    \includegraphics[width=0.8\linewidth]{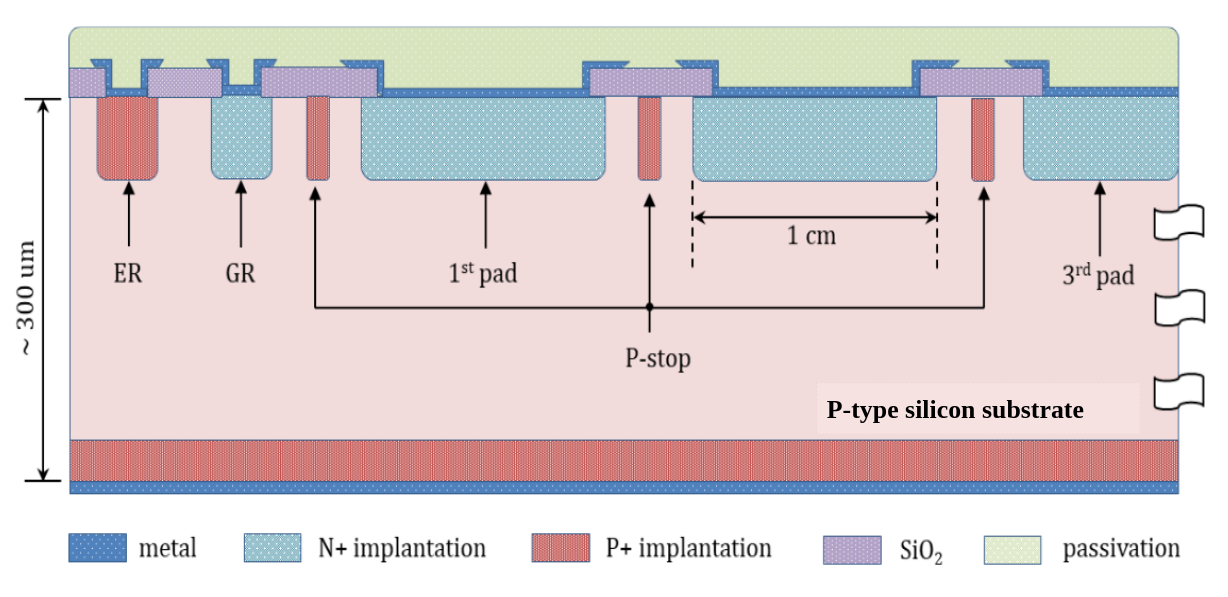}
    \caption{Simplified cross-section of the p-type pad detector}
    \label{sensor_design}
\end{figure}

Design optimisation was carried out with respect to key parameters such as 
doping concentration, pad geometry, inter-pad isolation, and guard ring 
configuration. Inter-pad isolation is necessary in n-in-p sensors because the 
positive oxide charge at the Si--SiO$_2$ interface causes electron accumulation 
that can short-circuit adjacent $n^{+}$ pads. This is mitigated using either 
p-stop (localised $p^{+}$ implants surrounding each $n^{+}$ pad) or p-spray 
(a uniform low-dose $p^{+}$ implantation across the full sensor surface), both 
of which suppress the surface electron accumulation layer. These optimisations are crucial for achieving high-voltage stability, minimising leakage current, and reducing junction capacitance, thereby improving the signal-to-noise ratio. Figure~\ref{sensor_design} shows a simplified cross-section of the p-type pad detector.

\subsection{Design Mask}

\begin{figure}[htbp]
    \centering
    \begin{subfigure}[b]{0.54\textwidth}
        \centering
        \includegraphics[width=\linewidth]{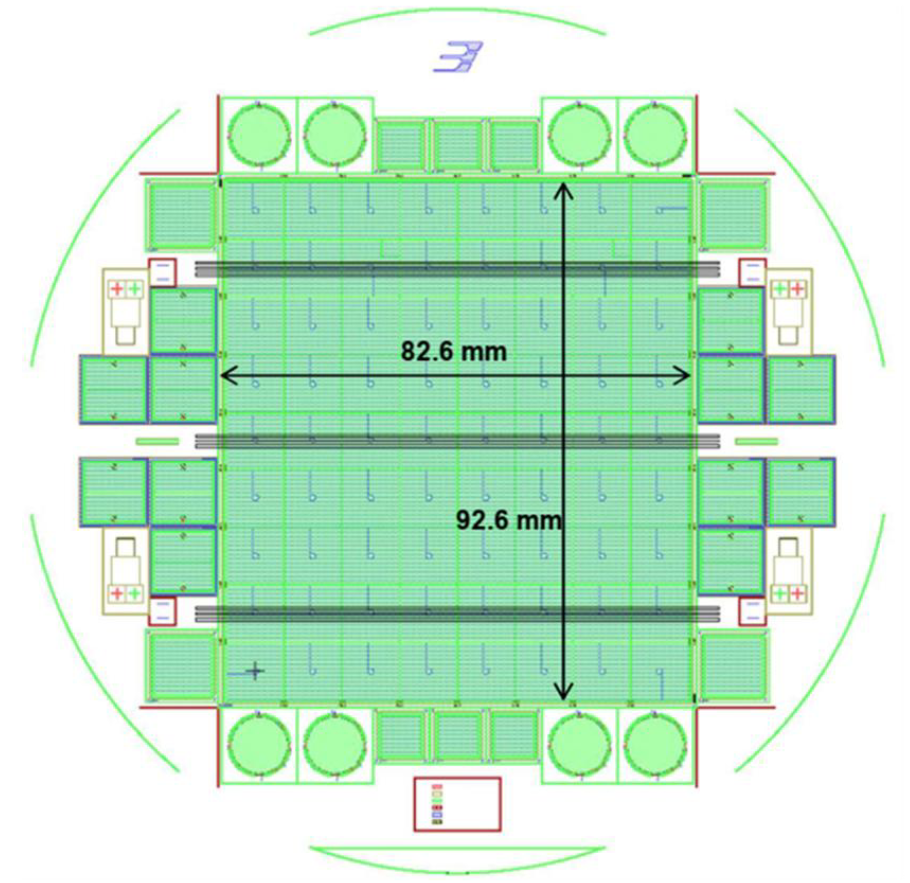}
        \caption{}
        \label{fig:mask_layout}
    \end{subfigure}
    \hfill
    \begin{subfigure}[b]{0.42\textwidth}
        \centering
        \includegraphics[width=\linewidth]{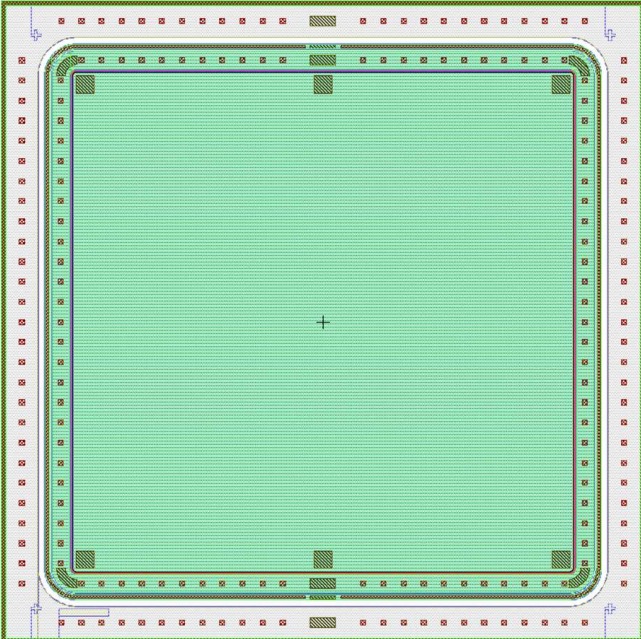}
        \caption{}
        \label{fig:test_structure}
    \end{subfigure}
    \caption{(a) Schematic of the final mask layout of the $8\times9$ pad detector 
    with peripheral test structures on a 6-inch wafer. 
    (b) Top view of a single-pad test structure diced from the peripheral region, 
    showing the active area, guard ring layout, and bond/probing pads.}
    \label{fig:sensor_design}
\end{figure}

The design employs six different mask sets, corresponding to $n^{+}$, $p^{+}$, 
p-stop, contact, metal, and passivation layers. The central region of the 6-inch wafer consists of an $8\times9$ array of silicon pad sensors, each with an active area of 1~cm$^2$. The peripheral region contains single-pad test structures (each with $1~\mathrm{cm}^2$ active area). These single-pads were diced from the wafer periphery and used for the irradiation studies reported in this work (see Figure~\ref{fig:sensor_design}). All test structures share 
identical geometry, guard ring configuration, and fabrication mask set as the 
central array sensors, differing only in their position on the wafer. It should 
be noted that these peripheral structures generally exhibit inferior performance 
compared to those located in the central region.

\section{Radiation exposure and defect assessment}
\label{sec:radiation_assessment}
 In high-luminosity high-energy physics experiments, surrounding detectors are exposed to intense particle fluence, leading to significant cumulative dose. The challenge gets manifold in the forward rapidity region especially due to significantly higher particle fluence, with neutrons often dominating the radiation spectrum around the beam direction. The reliable operation of sensors in these regions demands careful design consideration during sensor development and qualification.

The radiation field in high-energy physics experiments comprises a mixture of 
charged particles, photons (including X-rays and gamma rays), and neutrons. 
Sustained exposure to this mixed field leads to bulk degradation in silicon sensors, 
arising primarily from displacement of lattice atoms due to interactions with hadrons 
and neutrons. This is quantified by the Non-Ionizing Energy Loss (NIEL), expressed 
in terms of 1~MeV neutron equivalent fluence per square centimeter 
($n_{\mathrm{eq}}$/cm$^2$) \cite{Srour2003,CernNIEL}. In forward regions, where 
neutrons constitute the majority of the radiation field, neutron-induced displacement 
damage becomes the primary concern \cite{ATLAS2019Modeling}. This work therefore 
focuses on bulk damage induced by neutron irradiation, relevant to the forward 
calorimetry application for which these sensors are designed.

\begin{figure*}%[h!]
    \centering
    \includegraphics[width=\textwidth]{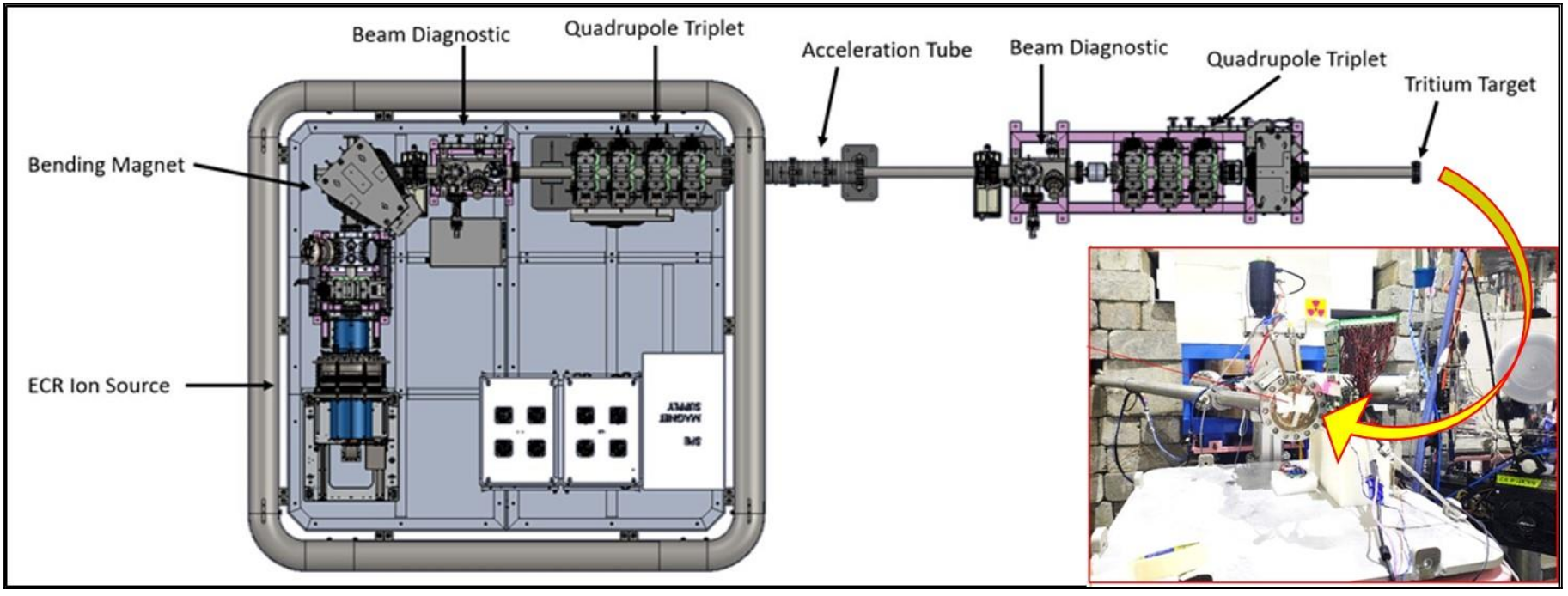}
    \caption{(Left) Schematic of the neutron irradiation facility at IPR. Actual placement of a test sensor during the irradiation test is shown in the inset figure.}
    \label{exp_setup}
\end{figure*}

 FLUKA simulations for the forward calorimeter of ALICE experiment suggest that the innermost silicon layers may experience a cumulative fluence on the order of $7\times10^{13}$~1~MeV~$n_{\mathrm{eq}}$/cm$^2$ over an operational lifetime of 10 years, including a safety factor of 10 \cite{ALICE_FoCal_TDR_2024}. We note that this is modest compared to the fluences of up to $\sim10^{16}$~$n_{\mathrm{eq}}$/cm$^2$ anticipated in the innermost layers of HL-LHC tracking and endcap calorimeter systems \cite{CMSHGCal, ATLASITk}; however, it represents the relevant design target for the low-luminosity heavy-ion forward calorimetry application motivating this sensor development.

\section{Irradiation technique and experimental setup}
\label{sec:setup}

Radiation damage studies were carried out at the Neutron and Ion Irradiation Facility operated by the Institute for Plasma Research (IPR), Ahmedabad. This facility employs an accelerator-based D–T neutron generator built for high-flux neutron production and controlled irradiation experiments \cite{Vala2025}. Neutrons are generated via the ${^3}$H(D,n)${^4}$He fusion reaction, wherein deuterium ions (D$^+$) are accelerated to energies of up to 300~keV using an electrostatic accelerator and impinged upon a tritiated titanium (TiT) target. The system can deliver an average neutron yield of approximately $1.2 \times 10^{12}$~n/s, with peak outputs reaching up to $5 \times 10^{12}$~n/s, making it well-suited for detailed radiation tolerance studies of semiconductor devices.

A schematic of the irradiation setup is shown in 
Figure~\ref{exp_setup}. The neutron fluence was determined by activation of indium foils wrapped directly around the sensors during irradiation, using the $^{115}$In(n,n')$^{115m}$In reaction with activation cross sections taken from the IAEA International Reactor Dosimetry and Fusion File (IRDFF-II)~\cite{IRDFF2020}, following the foil activation method of Ref.~\cite{Kim1989}. The raw neutron fluence was converted to 1~MeV neutron-equivalent fluence using the hardness factor $\kappa = 1.88$ for 14.1$\sim$ MeV D--T neutrons~\cite{Fretwurst1993}. The total uncertainty on the quoted fluence values, arising from the activation cross-section and counting statistics, is estimated to be 5\%~\cite{Fretwurst1993}.

For this study, five single-pad silicon test  sensors were selected for irradiation. One sensor served as a non-irradiated reference, while the remaining four were subjected to neutron fluences of approximately $10^7$, $10^{10}$, $10^{13}$, and $10^{14}$ $n_{eq}$/cm$^2$, as summarised in Table~\ref{sensor_dose}. These selected dose levels were chosen to encompass a broad range of operational scenarios, including the projected target fluence of $\sim7\times10^{13}$~1~MeV~
$n_{\mathrm{eq}}$/cm$^2$ for the forward calorimetry application 
motivating this sensor development \cite{ALICE_FoCal_TDR_2024}. By irradiating different sensors at increasing dose steps, rather than subjecting a single device to a cumulative dose, the study enables independent characterisation 
of leakage current evolution and annealing behaviour at each distinct fluence level, 
without the confounding effects of accumulated thermal history.

\begin{table}[h!]
\caption{Measured 1\,MeV neutron-equivalent fluence for each sensor,
listed in order of increasing fluence. Sensor labels D0--D4 are
assigned in order of increasing fluence.
The quoted uncertainty of 5\% arises from the activation
cross-section and counting statistics of the
$^{115}$In(n,n')$^{115m}$In dosimetry~\cite{Fretwurst1993}.}
\label{sensor_dose}
\centering
\begin{tabular}{|p{2.5cm}| p{3cm} | p{6cm} |}
\hline
Sensor label  & Accumulated fluence (1\,MeV~$n_{\mathrm{eq}}$/cm$^{2}$) \\
\hline\hline
D0  & 0 \quad (non-irradiated reference) \\
D1  & $(4.9 \pm 0.2) \times 10^{7}$ \\
D2  & $(1.1 \pm 0.1) \times 10^{10}$ \\
D3  & $(5.0 \pm 0.3) \times 10^{13}$ \\
D4  & $(2.5 \pm 0.1) \times 10^{14}$ \\
\hline
\end{tabular}
\end{table}

The current–voltage (I–V) characteristics of the irradiated silicon sensors were systematically measured at regular intervals following exposure. Radiation-induced displacement damage, primarily due to energetic neutron interactions, is known to introduce crystal lattice defects, which in turn elevate the bulk leakage current.

To assess the post-irradiation charge collection performance, the sensors were tested using a standard $^{90}Sr$  $\beta$ source, whose beta spectrum extends to an endpoint energy of 2.28 MeV; these electrons serve as minimum ionizing particles (MIP)  for silicon sensor characterisation. The response of the irradiated sensors was compared against that of a non-irradiated reference sensor.

\section{Results and Discussion}

\begin{figure}
    \centering
     \includegraphics[height=0.4\textheight,keepaspectratio]{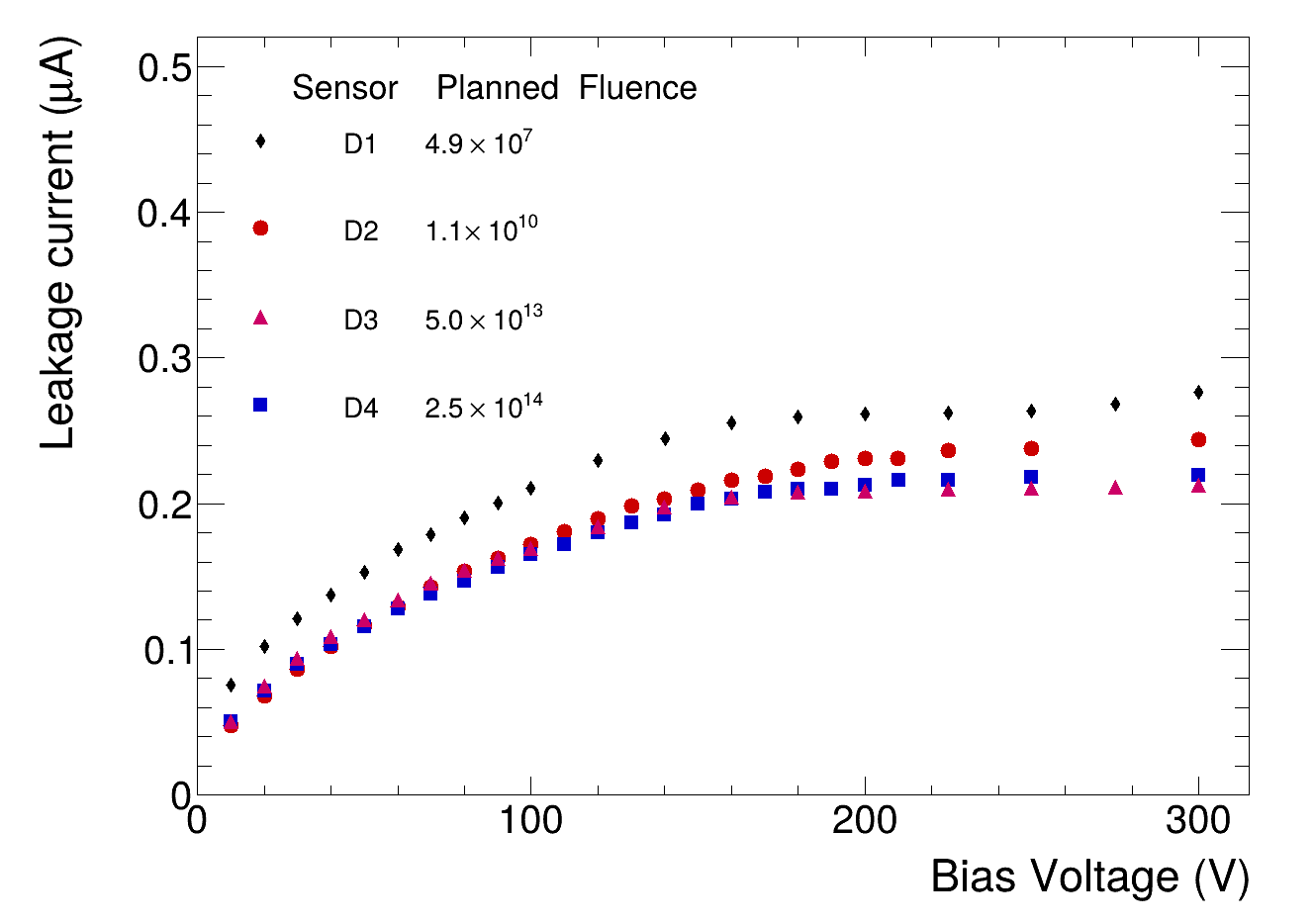}
    \caption{Individual leakage current as a function of  reverse bias voltage (I-V) before irradiation.}
    \label{IV_results_before_irradiation}
\end{figure}

Figure~\ref{IV_results_before_irradiation} presents the leakage current characteristics of all test sensors prior to irradiation. The measured leakage currents were consistently within 250 $nA$ across the set, and exhibited uniform behavior as a function of applied reverse bias. All four non-irradiated sensors displayed similar I–V trends, with the leakage current saturating near 160~V, indicating that full depletion was achieved around this bias level.

\begin{figure*}[htbp]
  \centering
  \includegraphics[width=\textwidth,height=0.5\textheight,keepaspectratio]{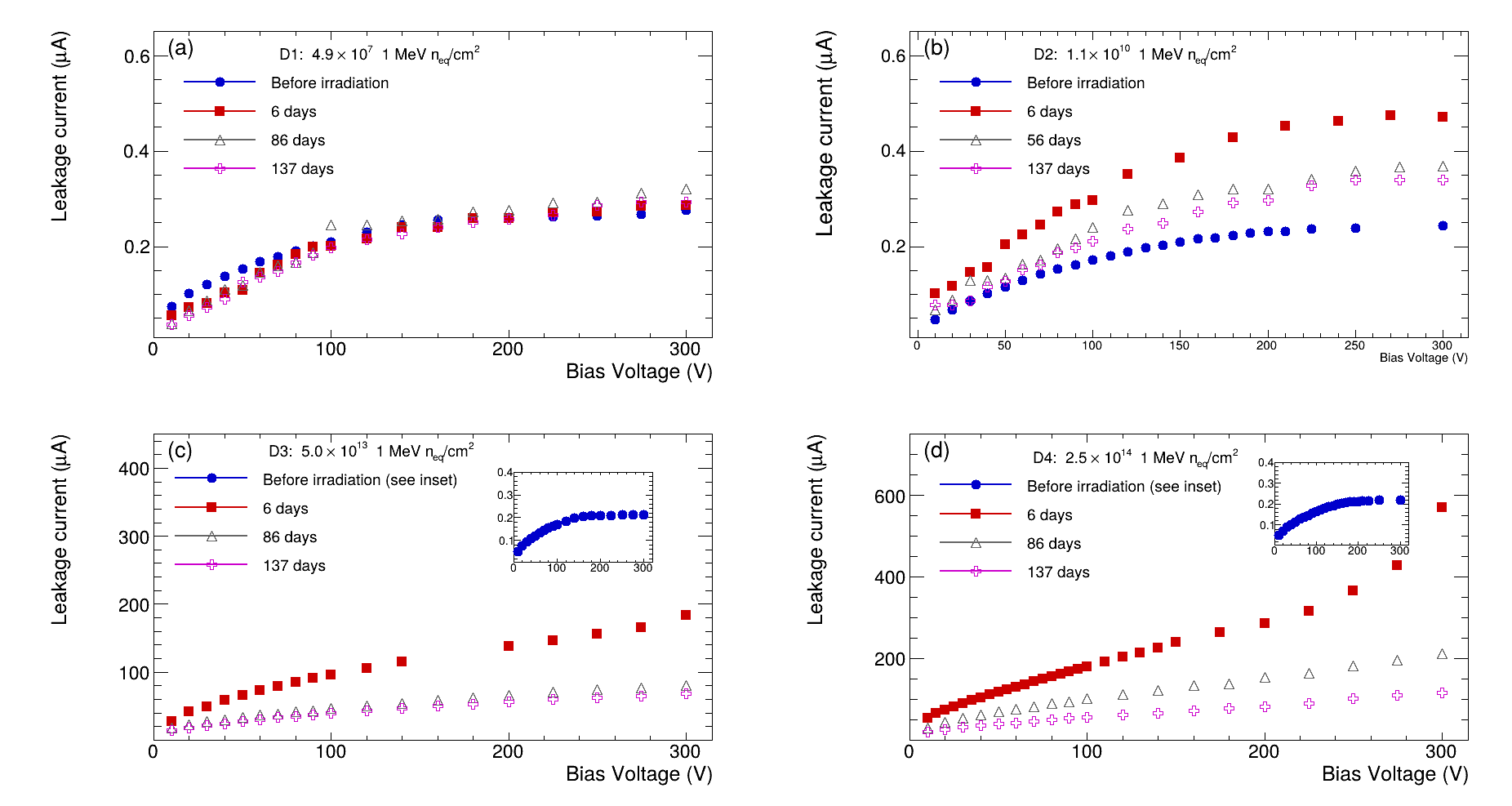}
  \caption{I-V characteristics after irradiation for different sensors. The number of days elapsed since irradiation are mentioned in the legend. For the sensor with higher irradiation dose, their current levels before irradiation are shown as insets with bias voltage measured in Volts(V) and leakage current measured in micro amperes ($\mu A$).}
  \label{IV_after_irradiation}
\end{figure*}

Among the sensors tested, the maximum leakage current observed was 0.28\,$\mu$A, recorded for sensor D1 at an applied voltage of 300~V. Owing to its comparatively high baseline performance, this sensor was selected for irradiation at the lowest dose level in the subsequent study.
Following neutron irradiation, all test sensors were subjected to a mandatory cooling-off period to allow residual activity to decay for safe handling. Subsequently, leakage current measurements were performed at regular intervals to monitor radiation-induced effects. During measurements, the cathode ($n^{+}$ pad) and the guard ring (also an $n^{+}$ 
implant) were kept at a positive bias relative to the anode ($p^{+}$ back contact). 
The bulk leakage current through the sensor pad and the surface leakage current 
through the guard ring were measured separately via independent channels of the 
power supply. All I--V measurements were performed at room temperature (20--25$^\circ$C). 
The sensors were irradiated at room temperature at the IPR neutron irradiation 
facility and allowed to cool for approximately 24 hours following irradiation 
before being transported for measurement. The time $t = 0$ is defined as the 
end of the irradiation exposure. Although measurements were initiated within 
6 days of irradiation, the leakage current of higher-dose sensors exhibited 
instability immediately after irradiation --- upon application of reverse bias, 
the current continued to rise rather than settling to a stable value. Stable and reproducible leakage current values were obtained from 6 days post-irradiation onwards, and all reported measurements correspond to this stable regime. The first data point in the annealing curves therefore corresponds to $t \approx 6$~days, and the single-exponential fit is applied 
over the full observation window of $\sim$140~days. It should be noted that 
all annealing reported in this work is purely natural, occurring at ambient 
room temperature without any deliberate thermal treatment. No accelerated 
annealing steps (such as heating to 60$^\circ$C as used in the RD48 
protocol) were applied at any stage. The evolution of leakage current for each irradiated sensor is shown in Figure~\ref{IV_after_irradiation}.

The sensor~(D1) exposed to a measured fluence of $(4.9 \pm 0.2) \times 10^{7}$~$n_{\mathrm{eq}}/\mathrm{cm}^2$
exhibited negligible radiation effects, with the post-irradiation leakage current 
remaining comparable to pre-irradiation levels. The sensor~(D2) irradiated at an 
intermediate fluence of $1.1 \times 10^{10}$~$n_{eq}/\text{cm}^2$ showed an 
approximately twofold increase in leakage current immediately after exposure, 
which diminished over time, consistent with beneficial annealing processes. 
Similarly, the sensor~(D3) irradiated at $(5.0 \pm 0.3) \times 10^{13}$~$n_{\mathrm{eq}}/\mathrm{cm}^2$
showed a significant increase in leakage current immediately after irradiation, 
with a gradual recovery observed over the measurement period. The sensor (D4) subjected to the highest fluence of $(2.5 \pm 0.1) × 10^{14}$~$n_{eq}/\text{cm}^2$ 
experienced pronounced degradation, with leakage current increasing by more than 
three orders of magnitude immediately after irradiation. Over a period of 
approximately two months, this current was observed to decrease by nearly 50\%, 
indicating the onset of long-term annealing effects. Despite the severity of the 
initial damage, all irradiated sensors demonstrated measurable, gradual recovery 
over time.

\begin{figure*}
    \centering
    \includegraphics[width=\textwidth]{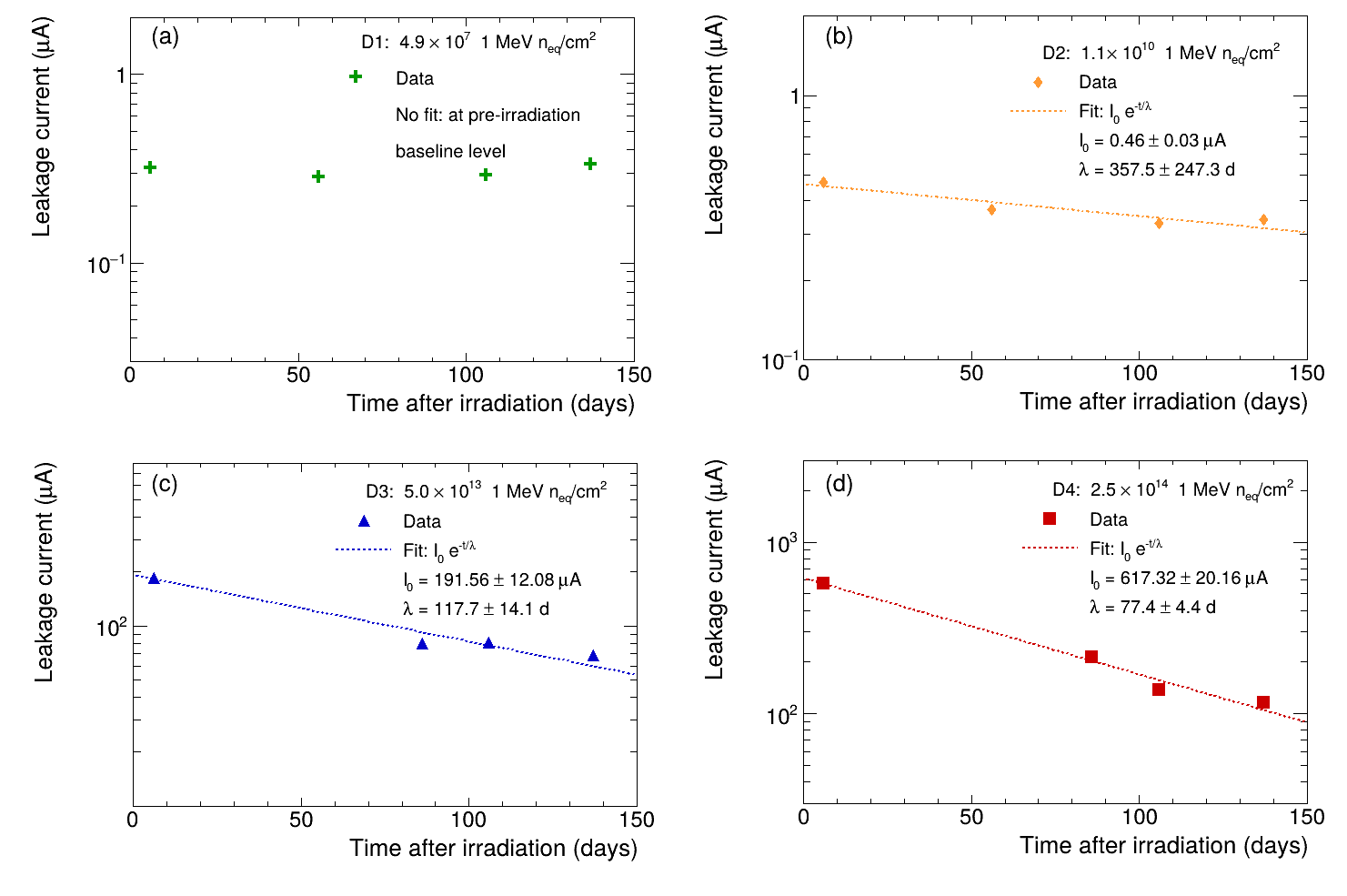}
    \caption{Leakage current (at 300 V) as a function of time after irradiation for the test sensors.}
    \label{current_vs_time_stacked}
\end{figure*}

%\subsection{Leakage-current recovery and dose scaling}
Figure~\ref{current_vs_time_stacked} shows the leakage current at a fixed bias of 
300 ~V as a function of time after irradiation for all sensors. The current drops 
rapidly immediately after irradiation and then decreases more slowly, consistent 
with the well-known beneficial annealing of radiation-induced bulk defects in 
silicon~\cite{Lindstrom2001}.

Radiation damage in silicon creates a spectrum of point defects and defect 
clusters --- primarily vacancies, interstitials, and their complexes with dopant 
atoms --- that introduce energy levels deep within the bandgap. These mid-gap 
states act as generation--recombination centres via the Shockley--Read--Hall 
mechanism, and it is the thermal generation of electron--hole pairs through these 
states that gives rise to the excess leakage current. Beneficial annealing refers 
to the thermally activated process by which a fraction of these defects migrate, 
recombine, or rearrange into electrically inactive configurations in the weeks and 
months following irradiation. Because the annealing rate is governed by the 
Arrhenius-like thermal activation of defect migration, it proceeds faster at room 
temperature than at cryogenic storage, and also faster in higher-dose sensors 
where the defect density --- and hence the probability of vacancy--interstitial 
recombination --- is larger, consistent with the dose-dependent recovery rate 
observed here.

\begin{figure*}[htbp]
    \centering
    \includegraphics[width=0.7\textwidth]{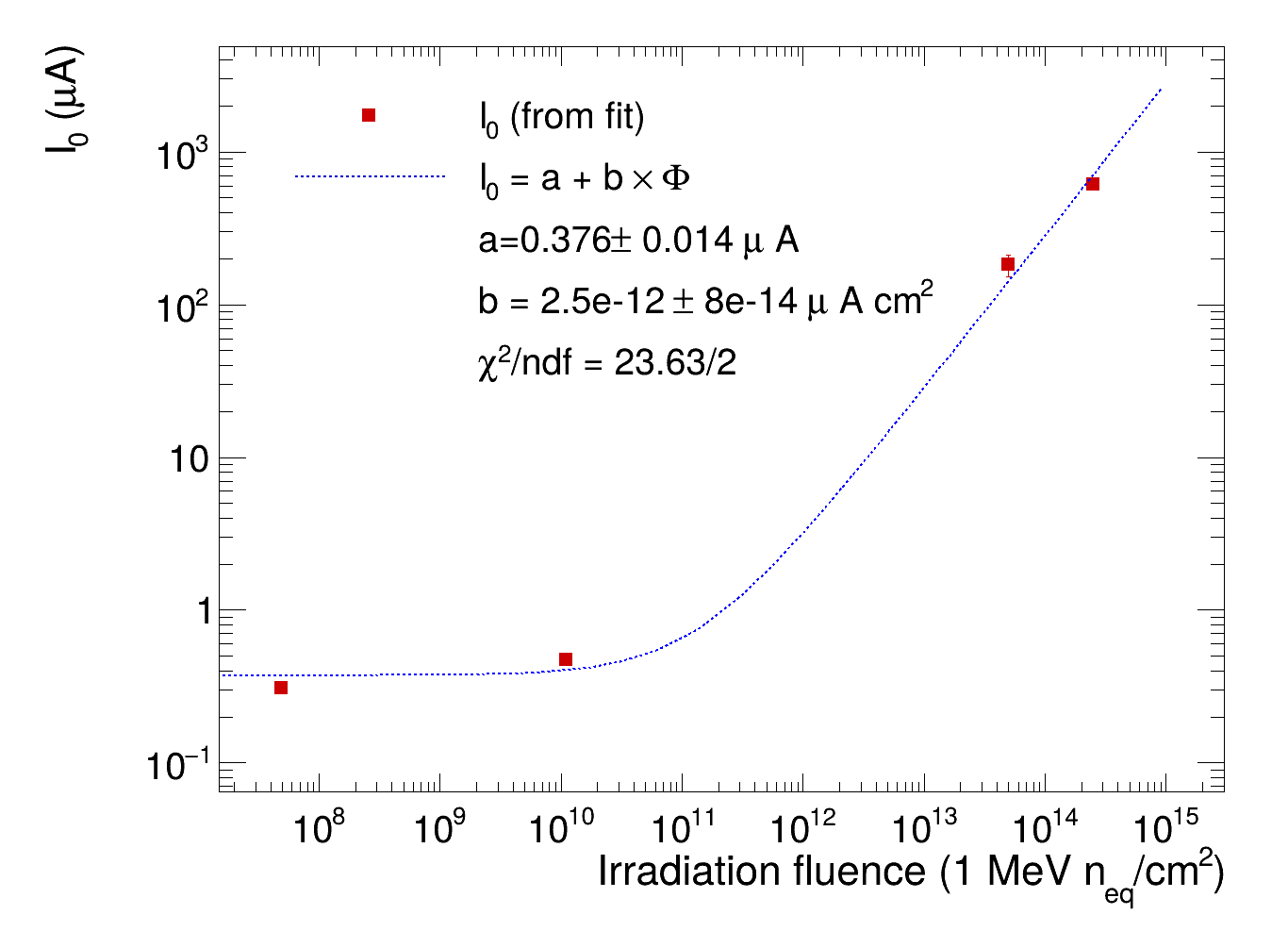}
    \caption{The maximum current(extrapolated from Figure~\ref{current_vs_time_stacked}) as a function of irradiation dose.}
    \label{I_0_vs_dose}
\end{figure*}

The recovery rate is observed to scale with the received dose --- higher-dose 
sensors exhibit a faster initial drop in leakage current. The complete description 
of leakage current annealing following the RD48 parameterisation includes both a 
sum of exponential components and a slowly-varying logarithmic term 
\cite{Moll1999}. However, within the $\sim$140-day observation window of this 
study, the measurement timescale is comparable to only the fast annealing 
component; the logarithmic term, which becomes significant only beyond a few 
multiples of $\lambda$, contributes negligibly over this range. A 
single-exponential form $I(t) = I_0\,e^{-t/\lambda}$ is therefore adopted as 
a physically motivated and statistically parsimonious description of the data 
within the observed time range. We verify that this form provides a good 
description of the data within the observation window; extrapolations beyond 
this range are explicitly flagged as lower bounds due to the expected onset of 
the logarithmic tail.  The physical origin of the logarithmic long-term behaviour is that 
radiation does not create a single type of damage with one characteristic recovery 
time; instead, it produces a wide variety of defects, each requiring a different 
amount of thermal energy to anneal away. The easiest defects to heal disappear 
first, giving the rapid initial drop in current. The remaining defects take  progressively longer and longer to recover. When many such 
processes with different recovery timescales are superimposed, the overall decay 
naturally becomes logarithmic rather than exponential --- the current continues to 
fall, but ever more slowly, with no single characteristic timescale. Unlike the effective doping concentration, which exhibits 
reverse annealing --- a long-term increase in depletion voltage driven by the 
transformation of certain defect complexes into more electrically active 
configurations --- the leakage current undergoes only beneficial annealing and 
decreases monotonically with time.
To quantify this behaviour, we fit the time evolution of each sensor with the 
single-exponential form defined above. The extracted $I_0$ values and their 
dependence on fluence are shown in Figure~\ref{I_0_vs_dose}, where a linear 
fit of the form $I_0(\Phi) = I_\text{dark} + k\,\Phi$ describes the data well, 
consistent with the standard relation $\Delta I/V = \alpha\,\Phi$ 
\cite{Moll1999, RD48Status3}. The active area used in this normalisation is the nominal geometric pad area, $A = 1~\mathrm{cm}^{2}$, consistent with standard practice for guarded pad diodes: the guard ring is designed to intercept edge and surface leakage current so that it does not contribute to the pad-channel current used here. The 60~$\mu$m gap between the pad edge and the guard-ring inner edge (Section~\ref{subsec:design_optimization}) is a region of ambiguous field-line behaviour; taking the area enclosed by the inner edge of the guard ring (pad $+$ gap, $1.012~\mathrm{cm}\times1.012~\mathrm{cm} = 1.024~\mathrm{cm}^{2}$) as a conservative upper bound on the true active area gives an upper-bound normalisation correction of $+2.4\%$, corresponding to a comparable downward revision of $\alpha$ if the entire gap were in fact collected by the pad. This is small compared to the $\sim$5\% fluence uncertainty (Table~\ref{sensor_dose}) and does not affect the conclusions of this work. The extracted current-related damage constant $\alpha \approx 8.3 \times 
10^{-17}$~A/cm is within a factor of two of the RD48 reference value of 
$(3.99 \pm 0.03) \times 10^{-17}$~A/cm \cite{RD48Status3}, which is defined 
after accelerated annealing at 60$^\circ$C for 80 minutes and normalised to 
20$^\circ$C. The comparison is not on strictly equal footing for the following 
reasons: (i) the sensors in this study underwent only natural room-temperature 
annealing over $\sim$140 days, without any accelerated thermal treatment; 
(ii) the annealing state at the time of $I_0$ extraction corresponds to the 
extrapolated current at $t=0$, not to a standardised annealing condition; 
and (iii) the hardness factor $\kappa = 1.88$ for 14.1~MeV D--T neutrons 
\cite{Fretwurst1993} was applied to convert the raw neutron fluence from indium 
foil activation to 1~MeV neutron-equivalent fluence. The observed factor-of-two 
discrepancy is consistent with the combined effect of the different annealing 
state and measurement conditions relative to the RD48 protocol.

\begin{figure*}[htbp]
    \centering
    \includegraphics[width=\textwidth]{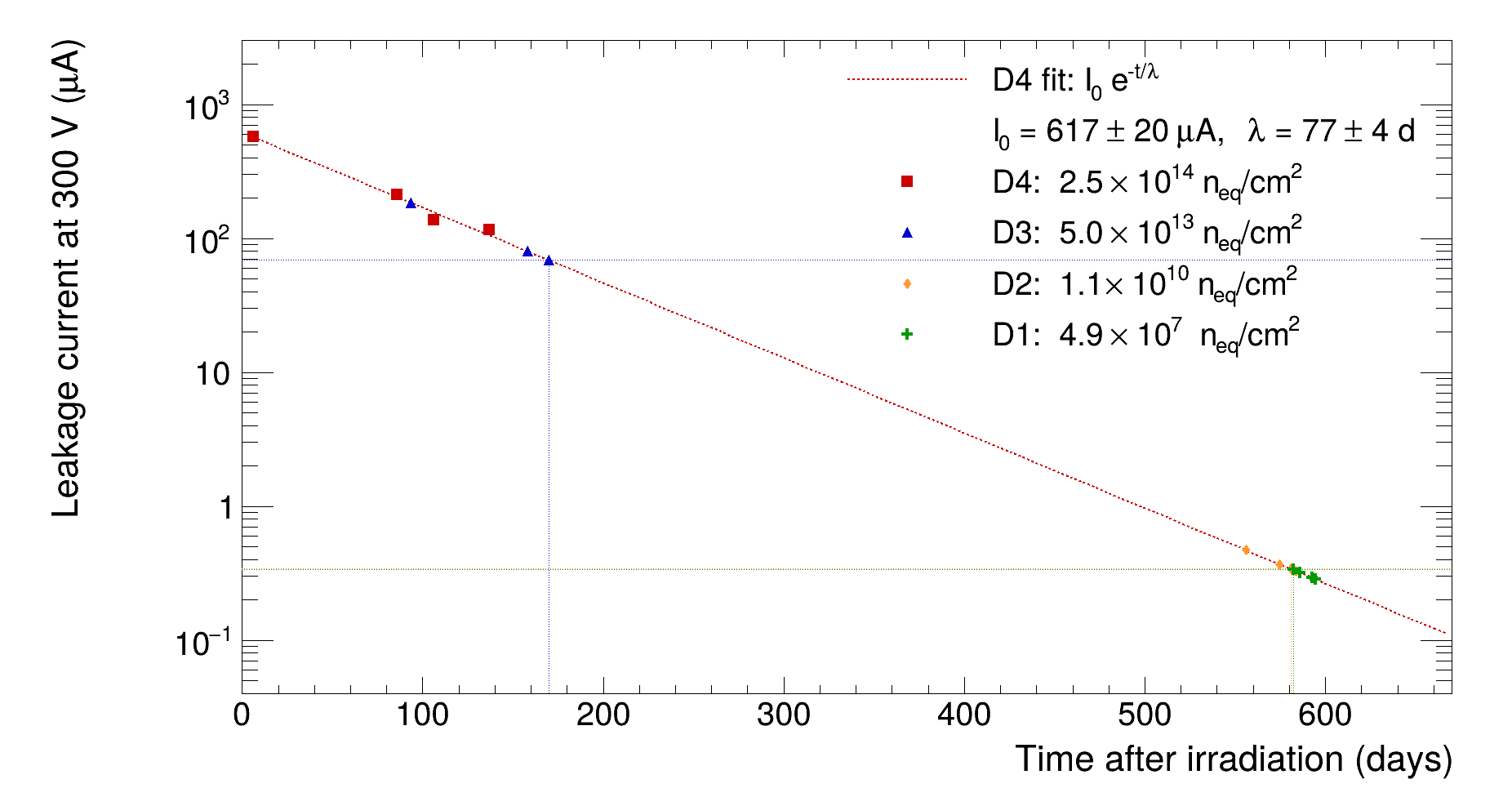}
    \caption{Leakage current at 300~V as a function of time after irradiation. The exponential fit for D4 has been extrapolated beyond the observation window. Data points for D1--D3 are overlaid at their projected equivalent times on the D4 curve, illustrating the estimated time for D4 to recover to the leakage current levels of lower-fluence sensors.}
    \label{leakage}
\end{figure*}

In Figure~\ref{leakage}, the exponential fit for the highest-dose sensor 
(D4, $\lambda = 77 \pm 4$~days) is extrapolated to longer timescales and overlaid 
with the current levels measured for lower-dose sensors, providing an estimate of 
the time required for D4 to recover to current levels characteristic of lower
fluences (Table~\ref{time_to_current}). 
 
The recovery time estimated for D3 (${\sim}80$~days) is reliable,
as it falls within a timescale comparable to the observation window, whereas the 
estimates for D2 and D1 (${\sim}460$ and ${\sim}500$~days, respectively) should
be regarded as optimistic lower bounds, since the logarithmic tail is expected to 
slow the recovery considerably at those timescales.

\begin{table}[htbp]
\caption{Estimated time for the leakage current of D4 to recover to
the level measured at 137 days for lower-fluence sensors. Estimates 
for D1 and D2 ($\sim$500 and $\sim$460 days respectively) should be 
regarded as optimistic lower bounds, as the logarithmic annealing tail 
is expected to slow recovery considerably beyond the observation window.}
\label{time_to_current}
\centering
\begin{tabular}{|p{2.5cm}| p{6cm} | p{2cm}|}
\hline
Sensor label & Accumulated fluence (1\,MeV~$n_{\mathrm{eq}}$/cm$^{2}$) & Time (days) \\
\hline\hline
D1 & $(4.9 \pm 0.2) \times 10^{7}$  & $\sim$500 \\
D2 & $(1.1 \pm 0.1) \times 10^{10}$ & $\sim$460 \\
D3 & $(5.0 \pm 0.3) \times 10^{13}$ & $\sim$80  \\
\hline
\end{tabular}
\end{table}

\begin{figure}[htbp]
    \centering
    \includegraphics[width=0.70\textwidth]{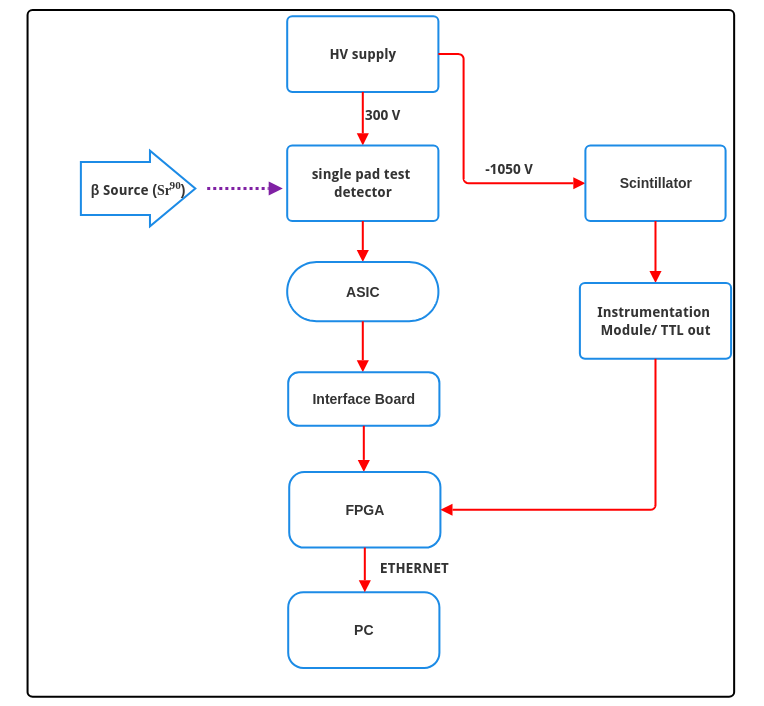}
    \caption{Schematic of the experimental setup for measuring the MIP signal 
    using a $^{90}$Sr $\beta$ source. The sensor is connected to a 
    16-channel mixed-signal front-end ASIC, which provides charge-sensitive 
    pre-amplification, pulse shaping, track-and-hold, and multiplexed readout. 
    An interface board provides bias voltage supply, signal routing, and 
    analogue-to-digital conversion for data acquisition.}
    \label{schematic_setup}
\end{figure}

\begin{figure}[htbp]
    \centering
    \includegraphics[width=\textwidth]{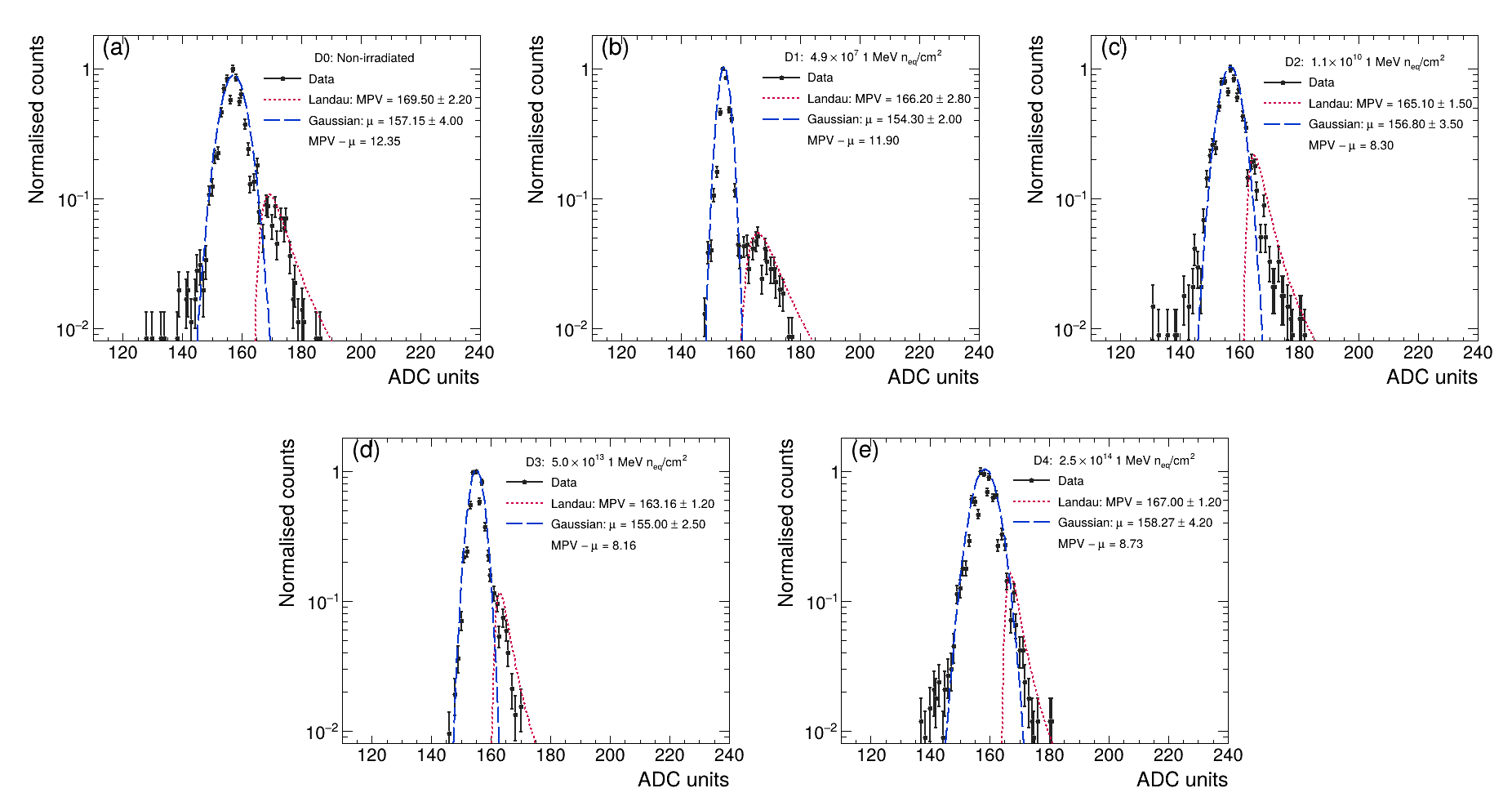}
    \caption{MIP detection response measured with a $^{90}\mathrm{Sr}$ $\beta$ source for sensors with different irradiation levels: (a)~D0, non-irradiated reference; (b)~D1, $(4.9 \pm 0.2) \times 10^{7}$~1\,MeV\,$n_{\mathrm{eq}}$/cm$^{2}$; (c)~D2, $(1.1 \pm 0.1) \times 10^{10}$~1\,MeV\,$n_{\mathrm{eq}}$/cm$^{2}$; (d)~D3, $(5.0 \pm 0.3) \times 10^{13}$~1\,MeV\,$n_{\mathrm{eq}}$/cm$^{2}$; (e)~D4, $(2.5 \pm 0.1) \times 10^{14}$~1\,MeV\,$n_{\mathrm{eq}}$/cm$^{2}$.}  
    \label{MIP_response}
\end{figure}

To assess the impact on calorimetric performance, the sensors were also tested 
with a $^{90}$Sr $\beta$ source, schematically shown in 
Figure~\ref{schematic_setup}. The measurements were taken 137 days after 
irradiation. For all sensors, the bias voltage was kept at 350~V. This bias 
was chosen because, as evident from the post-irradiation I--V characteristics 
at 137 days (Figure~\ref{IV_after_irradiation}), the leakage current saturates 
beyond $\sim$250~V for all sensors, indicating that 350~V places all sensors 
in the same saturation regime and allows a consistent comparison of MIP response 
across different irradiation levels. The leakage current values at 350~V for 
each sensor at the time of MIP measurement are summarised in 
Table~\ref{Current_at_350V}. While an unambiguous determination of the 
full-depletion voltage requires dedicated C--V measurements, which were not 
available in this study, the saturation of leakage current beyond 250~V 
suggests that all sensors were operating in a sufficiently depleted state at 
350~V. Since each sensor was measured independently, the absolute positions 
of the Landau MPV and Gaussian pedestal mean are not directly comparable 
across sensors due to possible differences in baseline settings. The 
physically meaningful quantity is the separation between the Landau MPV and 
the Gaussian pedestal mean, which reflects the signal-to-noise ratio. This 
separation is observed to decrease with increasing fluence, consistent with 
radiation-induced charge trapping reducing the collected charge and increasing 
leakage current broadening the noise baseline. The MIP response for all 
sensors is shown in Figure~\ref{MIP_response}.

\begin{table}[h!]
\caption{Measured 1\,MeV neutron-equivalent fluence for each sensor 
and leakage current at 300~V measured 137 days after irradiation. 
Since the leakage current saturates beyond $\sim$250~V for all 
sensors at this stage (see Figure~\ref{IV_after_irradiation}), the 
value at 300~V is equal to that at 350~V to within measurement 
precision.}
\label{Current_at_350V}
\centering
\begin{tabular}{|p{2.5cm}| p{4cm} | p{4cm}|}
\hline
Sensor label & Accumulated fluence (1\,MeV~$n_{\mathrm{eq}}$/cm$^{2}$) &
Leakage current at 300~V, 137 days ($\mu$A) \\
\hline\hline
D0 & 0 (non-irradiated) & $\approx 0.22$ \\
D1 & $(4.9 \pm 0.2) \times 10^{7}$  & 0.293 \\
D2 & $(1.1 \pm 0.1) \times 10^{10}$ & 0.340 \\
D3 & $(5.0 \pm 0.3) \times 10^{13}$ & 68.73 \\
D4 & $(2.5 \pm 0.1) \times 10^{14}$ & 116.57 \\
\hline
\end{tabular}
\end{table}

\section{Summary}
The p-type silicon pad sensors designed by BARC-VECC and fabricated at 
Bharat Electronics Limited (BEL), India, were developed to meet the 
requirements of forward calorimetry in high-luminosity heavy-ion collider 
experiments, with key design targets including a breakdown voltage exceeding 
1.2~kV, full depletion voltage of approximately 120~V, pad capacitance of 
$\sim$40~pF/cm$^2$, leakage current of the order of a few hundred nA at 
pre-irradiation, and operational integrity up to a cumulative fluence of 
$\sim7\times10^{13}$~1~MeV~$n_{\mathrm{eq}}$/cm$^2$ \cite{ALICE_FoCal_TDR_2024}.

Single-pad test structures diced from the peripheral region of the 6-inch 
wafer were irradiated with neutrons over a range of fluences from 
$\sim4.9\times10^{7}$ to $\sim2.5\times10^{14}$~1~MeV~$n_{\mathrm{eq}}$/cm$^2$. 
The leakage current was systematically monitored as a function of reverse 
bias voltage at regular intervals following irradiation. For fluences up to 
$\sim10^{10}$~$n_{\mathrm{eq}}$/cm$^2$, the leakage current remained within 
a few hundred nanoamperes, consistent with the pre-irradiation design target. 
At the target fluence of $\sim7\times10^{13}$~$n_{\mathrm{eq}}$/cm$^2$, the 
leakage current increased significantly but showed clear beneficial annealing 
over time. For sensors exposed to the highest fluence of 
$\sim2.5\times10^{14}$~$n_{\mathrm{eq}}$/cm$^2$, exceeding the design target 
by a factor of approximately three, the leakage current increased by more 
than three orders of magnitude immediately after irradiation but recovered 
by $\sim$50\% over approximately two months, demonstrating the resilience 
of the sensor and the fabrication process under conditions beyond the 
design requirement.

A single-exponential annealing model was introduced to describe the 
time-dependent leakage current recovery within the $\sim$140-day observation 
window, yielding a current-related damage constant $\alpha \approx 8.3\times
10^{-17}$~A/cm within a factor of two of the RD48 reference value 
\cite{RD48Status3}. The MIP response of the irradiated sensors was also 
studied, with the signal-to-noise separation decreasing with increasing 
fluence, consistent with radiation-induced charge trapping and increased 
noise from leakage current. Taken together, these results demonstrate that 
the indigenously fabricated sensors survive the target fluence with 
measurable but recoverable degradation, validating the BEL fabrication 
process and providing a baseline for future qualification of this domestic 
sensor production chain.

\acknowledgments

We thank the Department of Atomic Energy (DAE), India and Variable Energy Cyclotron Centre(VECC),Kolkata for providing the facilities and financial support for this work. We are grateful to Mr. Ranjay Laha and Mr. Arijit Das of Bharat Electronics Limited (BEL) for the timely delivery of the sensors. We also acknowledge the Institute for Plasma Research (IPR), Ahmedabad, for providing access to their neutron irradiation facility. Our thanks extend to the Head, Experimental High Energy Physics $\&$ Applications Group, VECC, for their support.

We gratefully acknowledge financial support from the DST–Government of India under the scheme Mega Facilities for Basic Science Research [Sanction Order No. SR/MF/PS-02/2021-Jadavpur (E-37128), dated December 31, 2021]. Finally, we thank Anamika Pallavi for her assistance in the laboratory during testing.

% Bibliography

%% [A] Recommended: using JHEP.bst file
%% \bibliographystyle{JHEP}
%% \bibliography{biblio.bib}

%% or
%% [B] Manual formatting (see below)
%% (i) We suggest to always provide author, title and journal data or doi:
%% in short all the informations that clearly identify a document.
%% (ii) please avoid comments such as "For a review'', "For some examples",
%% "and references therein" or move them in the text. In general, please leave only references in the bibliography and move all
%% accessory text in footnotes.
%% (iii) Also, please have only one work for each \bibitem.

\end{document}